\documentclass[%
 longbibliography,
 prapplied,%
 amsmath,amssymb,
 reprint,%
 superscriptaddress
,floatfix
]{revtex4-2}

\usepackage{graphicx}
\usepackage{siunitx}
\usepackage[utf8]{inputenc}
\usepackage{amsmath}
\usepackage{amsfonts}
\usepackage{amssymb}
\usepackage{natbib}
\usepackage{xcolor}
\usepackage{graphicx} 
\graphicspath{{figures/}}

\begin{document}

\title{Ge hole spin control using acoustic waves}
\date{\today}

\author{Chun-Yang Yuan}
\affiliation{Department of Physics, National Tsing Hua University, Taiwan}
\author{Tzu-Kan Hsiao}
\affiliation{Department of Physics, National Tsing Hua University, Taiwan}

\begin{abstract}
Germanium hole spin qubits based on strained Ge/SiGe quantum well have attracted much research attention due to the strong spin-orbit coupling. In particular, the strain dependence of the heavy-hole--light-hole mixing and thus the $g$-tensor anisotropy offer unique opportunities for acoustic driving and spin-phonon coupling. In this work we numerically simulate the coherent control of a Ge hole spin using surface acoustic waves. The periodic strain dynamically modulates the $g$-tensor matrix and causes fast spin rotation under a small acoustic amplitude. Moreover, we show a strong anisotropy and confinement dependence of the Rabi frequency coming from the phase-shifted longitudinal and shear strain components. Our work lays the foundations for acoustic-driven spin control and spin-phonon coupling using Ge hole spin qubits.

\end{abstract}

\maketitle


Hole spin qubits in strained Ge/SiGe quantum wells have rapidly emerged as a leading platform for quantum information processing~\cite{scappucci2021}. The high-quality heterostructures~\cite{Lodari2021, Stehouwer2023} and naturally abundant nuclear spin-free isotopes~\cite{Stehouwer2023} reduce the electric and magnetic disorder of the spin environment. The light effective mass of holes in Ge relaxes the stringent constraints on fabrication precision. Moreover, intrinsic spin-orbit coupling, which hybridizes heavy holes (HH) and light holes (LH), enables fast all-electrical spin control~\cite{Watzinger2018, Liu2023} and coherence sweep-spot operations~\cite{Hendrickx2024}. These advantages have facilitated the realization of a ten-spin qubit array~\cite{John2025}, an eight-site programmable quantum simulator~\cite{Farina2025}, universal control of four ST qubits~\cite{Zhang2025}, high-fidelity qubit gates~\cite{Lawrie2023, Wang2024}, and strong hole-photon coupling~\cite{DePalma2024}.

Notably, the strain dependence of the HH-LH mixing not only gives qubit tunability, but also opens new possibilities for acoustic spin control and spin-phonon coupling. Because of the strong dynamic strain modulation near the material surfaces, surface acoustic waves (SAWs) may offer a new control and coupling mechanism for hole spins confined in quantum wells. In quantum technologies, SAWs have been investigated as a vehicle for long-range qubit shuttling~\cite{McNeil2011, Hermelin2011, Jadot2021} or as a coupler to superconducting circuits~\cite{Gustafsson2014a, Satzinger2018a}. More recently, integration of hole spin qubits with SAWs has been proposed to achieve spin-phonon hybrid architectures~\cite{mei2025, KarimiYonjali2025}. However, so far the detailed mechanism of SAW-induced spin control and the resulting spin dynamics under acoustic drive remain unexplored.

\begin{figure}[b!]
    \centering
    \includegraphics[width=0.9\columnwidth]{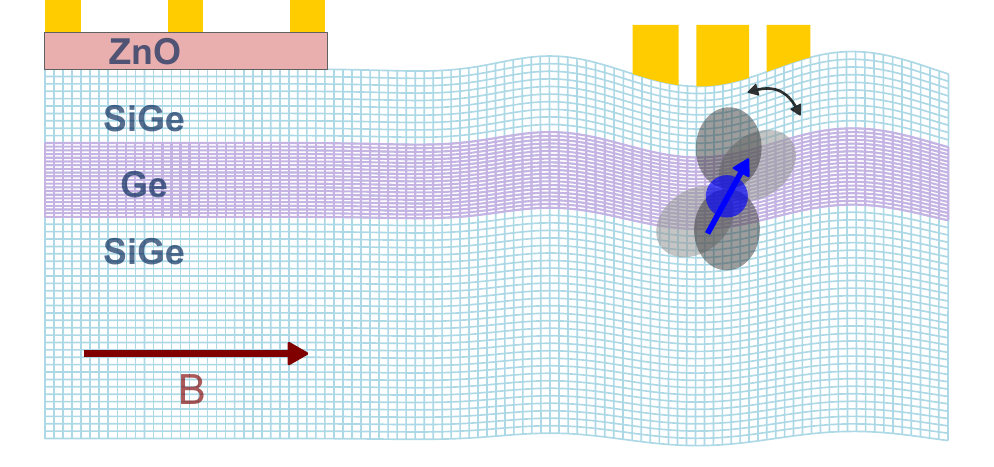} 
    \caption{\textbf{Device architecture.}
Schematic of the Ge hole spin qubit driven by a surface acoustic wave. An interdigital transducer (IDT), fabricated on a  ZnO film, generates a surface acoustic wave (SAW) propagating along the $[100]$ direction. The strain modulates the LH-HH mixing, causing the oscillating spatial profile of the $g$-tensor (represented by the shaking peanut shape).}
    \label{fig:device}
\end{figure}

In this letter, we investigate the mechanism governing acoustic-driven hole spin control in a realistic Ge/SiGe heterostructure device integrated with a SAW transducer, as illustrated in Fig.~\ref{fig:device}. First, we establish the theoretical framework by combining lattice elasticity theory with the Luttinger-Kohn and Bir-Pikus (LKBP) Hamiltonian to model the strain-induced g-tensor modulation. Next, we present numerical simulations of the SAW-driven Rabi oscillation, uncovering a distinct angular dependence that arises from the phase-shifted strain components. We then analyze this behavior geometrically, showing that the control efficiency is set by the relative rotation between the effective driving vector and the Larmor precession. Finally, we discuss the dependence of the acoustic driving efficiency on the dot geometry.



We consider a gate-defined hole spin qubit confined in a strained Ge/SiGe quantum well grown along the $z \parallel [001]$ direction. An interdigital transducer on a deposited piezoelectric ZnO film launches propagating surface acoustic waves which modulate the strain near the substrate surface~\cite{Mitsumoto2014}. The system is described by the $4 \times 4$ LKBP Hamiltonian $H_{LKBP}$ and the Zeeman Hamiltonian $H_Z$ in the basis of angular momentum states $|J, m_j\rangle$:
\begin{equation}
    H = \begin{pmatrix}
    P+Q & -S & R & 0 \\
    -S^\dagger & P-Q & 0 & R \\
    R^\dagger & 0 & P-Q & S \\
    0 & R^\dagger & S^\dagger & P+Q
    \end{pmatrix} + H_Z.
    \label{eq:LK_matrix}
\end{equation}
Here, the matrix elements contain both kinetic and strain contributions~\cite{Winkler2003} (see Appendix for the explicit form of the Hamiltonian). Unless indicated otherwise, we model the spatial confinement using a hard-wall potential with lateral dimensions $L_x = L_y = 50$~nm and a quantum well width $L_z = 15$~nm. We assume static strain due to lattice mismatch between the Ge quantum well and the SiGe layer $\epsilon_{xx} = \epsilon_{yy} = -0.61\%$ and $\epsilon_{zz} = 0.45\%$~\cite{Abadillo-Uriel2023}, which neglect the inhomogeneous strain fields typically induced by the metal gate stack~\cite{Frink2025}. Within the ground-state subband, the momentum operators are replaced by their expectation values: $\langle k_i^2 \rangle = (\pi/L_i)^2$ for $i \in \{x,y,z\}$~\cite{Venitucci2019}.

In the absence of an applied time-dependent electric field, the expectation values of the momentum operators remain zero ($\langle k_i \rangle = 0$) and by the assumption of square lateral confinement ($\langle k_x^2-k_y^2\rangle=0$), the kinetic mixing terms do not contribute to the driving dynamics. The coupling between heavy-hole ($m_j = \pm 3/2$) and light-hole ($m_j = \pm 1/2$) bands is therefore mediated exclusively by the time-dependent strain components entering the off-diagonal $S$ and $R$ terms.

A SAW with a frequency of 1~GHz propagates along the $x \parallel [100]$ crystal axis. We treat the SAW as a Rayleigh wave, where the displacement field $\mathbf{u}(\mathbf{r}, t)$ is characterized by coupled longitudinal ($u_x$) and transverse ($u_z$) motion:
\begin{equation}
\begin{aligned}
    u_x &= U e^{-k q_z z} \cos(kx - \omega t), \\
    u_z &= \eta U e^{-k q_z z} \cos(kx - \omega t + \delta), \\
    \epsilon_{ij} &= \frac{1}{2} \left( \frac{\partial u_i}{\partial x_j} + \frac{\partial u_j}{\partial x_i} \right).
\end{aligned}
\label{eq:displacement_strain}
\end{equation}
Here, $U$ represents the displacement amplitude at the surface ($z=0$), and the term $e^{-k q_z z}$ describes the exponential decay of the SAW mode into the substrate. Numerical solutions for our device configuration yield a specific amplitude ratio $\eta = 1.19$, a phase difference $\delta = 158^\circ$, and a decay constant $q_z = 0.44$~\cite{Mitsumoto2014}. The strain is evaluated at $z=60$~nm below the surface.
To capture the realistic heterostructure environment, we distinguish between the acoustic and electronic properties. The SAW propagation is modeled using the elastic constants of Si$_{0.2}$Ge$_{0.8}$, yielding a phase velocity $v \approx 3185$~m/s and a wavelength $\lambda \approx 3.19$~$\mu$m. We neglect the 15~nm Ge quantum well layer in the acoustic simulation since the thickness of the quantum well is much smaller than the wavelength. The system falls into the perturbation regime where the mode profile is dominated by Si$_{0.2}$Ge$_{0.8}$ substrate . In contrast, the hole spin dynamics within the quantum well are described using the Luttinger-Kohn parameters of stressed pure Ge.

We fix the displacement amplitude $U = 25$~pm at the surface unless stated otherwise. This value corresponds to an experimentally achievable displacement using a moderate RF power of $\sim -35$~dBm~\cite{Tucoulou2005}. Such a small amplitude ensures the system remains well within the linear regime, thereby validating the perturbation approximation

By projecting the full Hamiltonian onto the ground-state qubit subspace, this complex strain dynamics translates into a time-dependent $g$-tensor modulation $\Delta \mathbf{g}(t)$, leading to the effective driving Hamiltonian:
\begin{equation}
    H_{\text{eff}}(t) \approx \frac{1}{2} \mu_B \boldsymbol{\sigma} \cdot [\mathbf{g}_0 + \Delta \mathbf{g}(t)] \cdot \mathbf{B}.
    \label{eq:effective_ham}
\end{equation}

We define the orientation of the external in-plane magnetic field $\mathbf{B} = (B \cos\phi, B \sin\phi, 0)^T$ using spherical coordinates, where $\phi$ represents the azimuthal angle relative to the SAW propagation direction ($x$).

\begin{figure}[t!]
    \centering
    \includegraphics[width=\columnwidth]{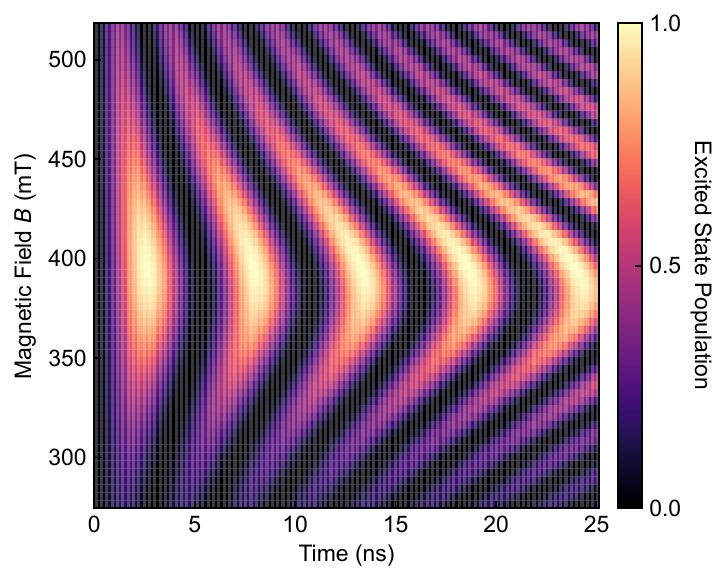}
    \caption{\textbf{SAW-driven spin control.}
        Time-resolved Rabi oscillation driven by the surface acoustic wave.
        The qubit is initialized in the ground state at $B = 0.396$ T, $\phi = 330^\circ$.
        The clear sinusoidal population transfer confirms coherent control via strain-induced $g$-tensor modulation.}
    \label{fig:single_rabi}
\end{figure}

To demonstrate acoustic spin control, we perform numerical simulations of the time-dependent $4 \times 4$ LKBP Hamiltonian using QuTip~\cite{Lambert2026}. Initially, the spin is prepared in the ground state without the SAW-induced dynamic strain. We then calculate the population of the first excited state as a function of the applied magnetic field magnitude $B$ and the SAW pulse duration. 

As shown in Fig.~\ref{fig:single_rabi}, for a magnetic field oriented at $\phi = 330^\circ$, the system exhibits clear coherent Rabi oscillations. Near the resonant magnetic field of $B = 0.396$~T, we extract a Rabi frequency of $\Omega_R/2\pi \approx 188.15$~MHz. This result confirms that the time-dependent strain generated by the SAW is a viable mechanism for fast hole spin driving in Ge quantum wells.

To get more insight into the driving mechanism, we numerically extract the time-dependent modulation of the $g$-tensor, $\Delta \mathbf{g}(t)$ (see Supplemental Material for details). Figure~\ref{fig:g_modulation} displays the evolution of the relevant tensor components. Due to the cubic crystal symmetry and the SAW propagation along the $[100]$ direction, the induced strain consists of longitudinal components ($\epsilon_{xx}, \epsilon_{zz}$) and a shear component ($\epsilon_{xz}$). Consequently, we observe modulations primarily in the in-plane diagonal elements ($\Delta g_{xx}, \Delta g_{yy}$) and the transverse coupling element ($\Delta g_{zx}$). components involving the transverse axis $y$  (e.g., $g_{xy}, g_{yz}$) remain static, and the modulation of $g_{zz}$ and $g_{xz}$ is found to be negligible due to the strong heavy-hole confinement along the growth direction.

Crucially, as observed in Fig.~\ref{fig:g_modulation}, there is a distinct phase shift between the modulation of the off-diagonal term $\Delta g_{zx}$ and the diagonal components ($\Delta g_{xx}, \Delta g_{yy}$). This result is in excellent agreement with the symmetry-based predictions for hole systems~\cite{Abadillo-Uriel2023}, where the $g$-tensor components are determined by specific strain-induced HH-LH mixing paths. Specifically, the transverse coupling $g_{zx}$ is primarily driven by the shear strain $\epsilon_{xz}$ entering the off-diagonal $S$ terms of the Hamiltonian. In contrast, the oscillations in the diagonal components $g_{xx}$ and $g_{yy}$ are modulated by the longitudinal strain $\epsilon_{xx}$ through its contribution to the $R$ terms. The phase shift between these $g$-tensor components thus directly reflects the intrinsic phase difference between the longitudinal and shear strain components of the Rayleigh-type SAW.

To understand the impact of this modulation on the spin qubit, we derive the effective driving vector $\mathbf{B_{eff}}(t)$ in the presence of a static in-plane magnetic field $\mathbf{B}$.  Using the dominant $\Delta \mathbf{g}(t)$ components identified above, $\mathbf{B_{eff}}(t)$ takes the form:
\begin{equation}
\begin{aligned}
    \mathbf{B_{eff}}(t) &\approx 
    \begin{pmatrix} 
        \Delta g_{xx}(t) & 0 & 0 \\ 
        0 & \Delta g_{yy}(t) & 0 \\ 
        \Delta g_{zx}(t) & 0 & 0 
    \end{pmatrix} \cdot 
    \begin{pmatrix}
         B  \cos\phi \\
         B  \sin\phi \\
         0
    \end{pmatrix} \\
    &= \begin{pmatrix}
        \Delta g_{xx}(t) B  \cos\phi \\
        \Delta g_{yy}(t) B \sin\phi \\
        \Delta g_{zx}(t) B \cos\phi
    \end{pmatrix}.
\end{aligned}
\label{eq:B_eff_matrix}
\end{equation}
This expression highlights how the strain-induced $g$-tensor components couple to the static magnetic field to generate a time-dependent driving vector. The Rabi frequency is determined by the component of $\mathbf{B_{eff}}(t)$ perpendicular to the static quantization axis.

\begin{figure}[t!]
    \centering
    \includegraphics[width=\columnwidth]{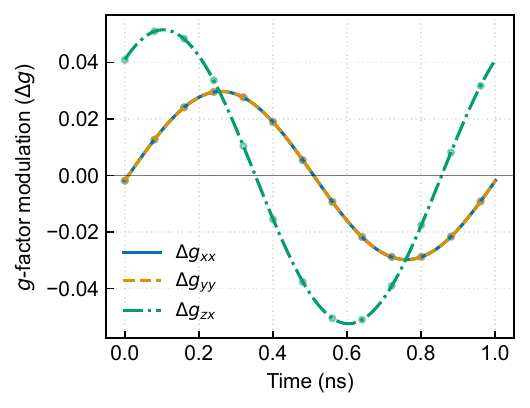} 
    \caption{\textbf{Time-dependent modulation of the $g$-tensor.}
    Calculated variation of the $g$-tensor components $\Delta g_{ij}(t)$ under the influence of the SAW strain over one acoustic period. The distinct phase shift between $\Delta g_{zx}$ and ($\Delta g_{xx}, \Delta g_{yy}$) is the origin of the elliptical polarized driving mechanism.}
    \label{fig:g_modulation}
\end{figure}

\begin{figure*}[t!]
    \centering
    \includegraphics[width=0.95\textwidth]{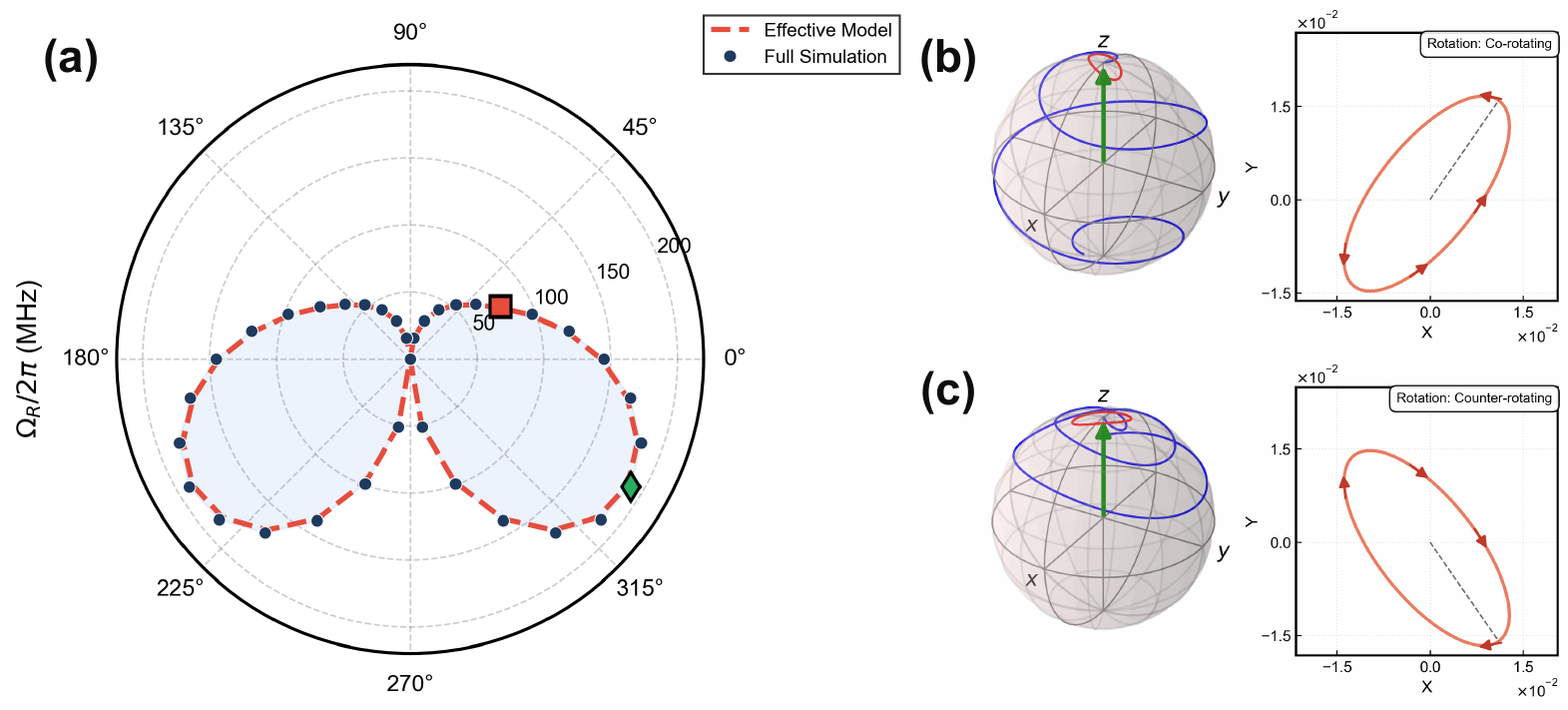} 
  \caption{\textbf{Anisotropy and chirality of SAW driving.} 
    (a) Rabi frequency $\Omega_R/2\pi$ as a function of the in-plane magnetic field angle $\phi$. The dots represent full numerical simulation results, while the dashed red line shows the excellent agreement achieved using the effective $g$-tensor modulation model.
    (b), (c) The left panel shows the trajectories of spin state (blue) and the effective driving vector (red) on the Bloch sphere evolved over the same time duration for $\phi = 330^\circ$ (green diamond in (a)) and $\phi = 30^\circ$ (red square in (a)), respectively. The right panel shows the ellipses projected on the x-y plane.
    The driving vector is elliptically polarized due to the intrinsic phase difference between the SAW strain components. At the optimal angle $\phi = 330^\circ$ (b), the effective vector co-rotates with the hole’s Larmor precession, maximizing the Rabi frequency. At $\phi = 30^\circ$ (c), it counter-rotates, suppressing the driving.}
    \label{fig:full_comparison}
\end{figure*}

We now examine the detailed dependence of the driving efficiency on the in-plane magnetic field angle $\phi$. We simulate the Rabi frequency $\Omega_R(\phi)$ by sweeping the field direction while maintaining the resonance condition. Figure~\ref{fig:full_comparison}(a) presents the results from the full LKBP and the effective $g$-tensor modulation model. The excellent agreement with the LKBP simulation confirms that the time-dependent $g$-tensor modulation captures the essential physics.
The angular dependence can be directly understood from Eq.~\eqref{eq:B_eff_matrix}. Specifically, the Rabi frequency drops to zero when the magnetic field is perpendicular to the SAW propagation direction ($\phi = 90^\circ$ or $270^\circ$). In these cases, Eq.~\eqref{eq:B_eff_matrix} reduces to $\mathbf{B_{eff}}(t) = (0, \Delta g_{yy}(t) B, 0)^T$, meaning the effective vector is purely along the $y$-axis, parallel to the static field $\mathbf{B}$. As a result, the Zeeman energy is modulated longitudinally with no transverse component to induce spin transitions, accounting for the observed nodes at these angles.
Beyond these nodes, a strong anisotropy is observed, with a striking lack of mirror symmetry with respect to the propagation axis ($\phi = 0$). This behavior cannot be explained by a simple linear driving, which would typically yield a symmetric response pattern. The origin of this asymmetry lies in the intrinsic properties of the Rayleigh wave. Due to the intrinsic phase difference $\delta \approx 158^\circ$ between the strain components, the transverse modulation $\mathbf{B_{eff}}(t)$ traces an elliptical trajectory rather than a linear oscillation.

To quantify the driving strength and explain the asymmetry, we analyze the transverse component of $\mathbf{B_{eff}}(t)$ in the reference frame of the spin. As detailed in the Supplemental Material, the Rabi frequency is determined by the geometric properties of this ellipse. Specifically, we derive that $\Omega_R \propto (R_{\text{major}} \pm R_{\text{minor}})$, where $R_{\text{major}}$ and $R_{\text{minor}}$ are the semi-major and semi-minor axes of the driving ellipse, and the sign is determined by the relative helicity between the driving vector and the spin Larmor precession.

We visualize this mechanism in Figs.~\ref{fig:full_comparison}(b) and (c). At the optimal angle $\phi = 330^\circ$, the effective vector co-rotates with the Larmor precession, satisfying the chirality matching condition. Here, the driving is dominated by the additive superposition of the circular components ($\Omega_R \propto R_{\text{major}} + R_{\text{minor}}$), resulting in a maximal Rabi frequency. Conversely, at $\phi = 30^\circ$, the vector is counter-rotating. This chirality mismatch indicates that the dominant component of the ellipse corresponds to the off-resonant term. The spin is thus driven only by the weak residual difference ($\Omega_R \propto R_{\text{major}} - R_{\text{minor}}$), explaining why the Rabi frequency is suppressed but remains non-zero.

In addition, we assess the controllability of the mechanism and its dependence on the dot geometry. Since the axes of the driving ellipse ($R_{\text{major}}, R_{\text{minor}}$) are directly proportional to the strain amplitude, our analytical model predicts a linear scaling of the Rabi frequency with the SAW displacement $U$.
\begin{figure}[b!]
    \centering
    \includegraphics[width=\columnwidth]{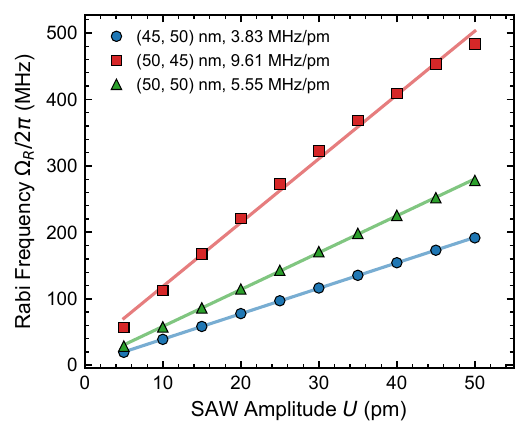} 
    \caption{\textbf{Rabi frequency dependence on the dot geometry.} 
    Calculated Rabi frequency $\Omega_R/2\pi$ as a function of the SAW displacement amplitude $U$ at $\phi = 0^\circ$ for various dot dimensions $(L_x, L_y)$. The solid lines represent linear fits, confirming that the driving mechanism is governed by first-order strain coupling. The variation in slopes illustrates the impact of dot geometry on driving efficiency.}
    \label{fig:power_dep}
\end{figure}
Figure~\ref{fig:power_dep} presents the numerical results at $\phi = 0^\circ$ for various dot dimensions $(L_x, L_y)$. We observe a strictly linear response ($\Omega_R \propto U$) for all configurations, consistent with the first-order dependence derived in the Supplemental Material.

The slope of the power dependence, representing the driving efficiency, varies with the dot aspect ratio. When the lateral symmetry is broken, the kinetic $R$ term in the Hamiltonian leads to a shift in the static $g$-factors. For the $(50, 45)$~nm configuration, the resulting reduction in $|g_{xx}|$ near the $x$-axis necessitates a larger external magnetic field $B$ to satisfy the resonance condition. Since the Rabi frequency is proportional to the resonant field magnitude ($\Omega_R \propto B$), this field enhancement leads to a significantly higher driving efficiency compared to the symmetric or $(45, 50)$~nm cases. More discussion about the dot geometry dependence is detailed in the Supplemental Material.



To conclude, by combining a realistic lattice elasticity model with the LKBP Hamiltonian, we have established that SAW-driven control of Ge hole spins is governed by an elliptical polarized driving vector $\mathbf{B_{eff}}(t)$ rooted in the inherent phase difference between longitudinal and transverse strain components of Rayleigh waves. The resulting Rabi frequency can theoretically exceed 100~MHz with an experimentally achievable SAW amplitude. Furthermore, the SAW-driven Rabi frequency exhibits a strong dependence on the alignment between the SAW propagating direction and the magnetic field as well as the dot geometry. These findings provide guidelines for acoustic-driven Ge hole spin qubit control.

Finally, we comment on strategies to further enhance spin-phonon coupling. Through strain engineering with unstrained Ge~\cite{Costa2025} or gate oxide~\cite{Frink2025}, combined with dot confinement, optimized field alignment, IDT design~\cite{KarimiYonjali2025} and phononic cavity~\cite{mei2025}, it becomes possible to access the strong coupling regime between a hole spin and a single SAW phonon, establishing new interconnections in quantum information processing.



\section*{Acknowledgment}
We acknowledge useful discussions with members of the Hsiao group. We thank the support from the National Science and Technology Council (NSTC) grants (112-2112-M-007-054-MY3) and (114-2119-M-007-008). We also acknowledge Yushan Fellow Program (MOE-111-YSFMS-0002-001-P1) and the Center for Quantum Science and Technology (CQST) by the Ministry of Education (MOE), Taiwan for the financial support.
\section*{Data Availability}
The data reported in this paper are archived on a Zenodo data repository at \url{https://doi.org/10.5281/zenodo.18079731}

\bibliography{SAW_paper_ref}

\appendix
\section{Explicit Form of the Hamiltonian and Basis States}
\label{app:hamiltonian}

In our numerical simulations, we model the hole spin dynamics using the $4 \times 4$ Luttinger-Kohn-Bir-Pikus Hamiltonian within the basis of angular momentum eigenstates $|J=3/2, m_J\rangle$, ordered as $\left\{ |+\frac{3}{2}\rangle, |+\frac{1}{2}\rangle, |-\frac{1}{2}\rangle , |-\frac{3}{2}\rangle\right\}$. The total Hamiltonian is given by $H = H_{LK} + H_{BP} + H_Z$.

The kinetic ($H_{LK}$) and strain ($H_{BP}$) contributions are combined into the following matrix structure:
\begin{equation}
    H_{LK} + H_{BP} = \begin{pmatrix}
    P+Q & -S & R & 0 \\
    -S^\dagger & P-Q & 0 & R \\
    R^\dagger & 0 & P-Q & S \\
    0 & R^\dagger & S^\dagger & P+Q
    \end{pmatrix},
\end{equation}
where the matrix elements are sums of kinetic ($K$) and strain ($\epsilon$) terms: $P = P_K + P_\epsilon$, $Q = Q_K + Q_\epsilon$, etc.
The kinetic terms are defined as:
\begin{equation}
\begin{aligned}
    P_K &= \frac{\hbar^2\gamma_1}{2m_0} (k_x^2 + k_y^2 + k_z^2), \\
    Q_K &= \frac{\hbar^2\gamma_2}{2m_0} (k_x^2 + k_y^2 - 2k_z^2), \\
    R_K &= \frac{\hbar^2\sqrt{3}}{2m_0} \left[-\gamma_2(k_x^2 - k_y^2) + 2i\gamma_3 k_x k_y\right], \\ 
    S_K &= \frac{\hbar^2\sqrt{3}}{m_0} \gamma_3 (k_x - ik_y)k_z,
\end{aligned}
\end{equation}
where $m_0$ is the free electron mass and $\gamma_{1,2,3}$ are the Luttinger parameters. Note that $S_K$ vanishes at the subband edge in symmetric wells.

The strain-induced terms, governed by the deformation potentials $a_v, b_v, d_v$, are given by:
\begin{equation}
\begin{aligned}
    P_\epsilon &= -a_v (\epsilon_{xx} + \epsilon_{yy} + \epsilon_{zz}), \\
    Q_\epsilon &= -\frac{b_v}{2} (\epsilon_{xx} + \epsilon_{yy} - 2\epsilon_{zz}), \\
    R_\epsilon &= \frac{\sqrt{3}}{2} b_v (\epsilon_{xx} - \epsilon_{yy}) - i d_v \epsilon_{xy}, \\
    S_\epsilon &= -d_v (\epsilon_{xz} - i \epsilon_{yz}).
\end{aligned}
\end{equation}
Crucially, the Rayleigh SAW introduces a shear strain component $\epsilon_{xz}$, which activates the $S_\epsilon$ term responsible for the heavy hole-light hole mixing and the resulting $g$-tensor modulation.

Finally, the Zeeman Hamiltonian accounts for both isotropic and anisotropic cubic contributions:
\begin{equation}
    H_Z = 2\mu_B \left[ \kappa \mathbf{J} \cdot \mathbf{b} + q (J_x^3 B_x + J_y^3 B_y + J_z^3 B_z) \right],
\end{equation}
where $\mu_B$ is the Bohr magneton, $\mathbf{J}$ represents the spin-3/2 matrices, and $\kappa, q$ are the magnetic Luttinger parameters.
The material parameters for Germanium used in our simulations are summarized in Table~\ref{tab:material_params}.

\begin{table}[h]
    \centering
    \caption{Germanium material parameters used in the simulations. Deformation potentials ($a_v, b_v, d_v$) are in eV; other parameters are dimensionless.}
    \label{tab:material_params}
    \setlength{\tabcolsep}{4pt} 
    \begin{tabular}{ccccccccc}
    \hline \hline
    $\gamma_1$ & $\gamma_2$ & $\gamma_3$ & $\kappa$ & $q$ &$a_v$(eV)& $b_v$(eV) & $d_v$(eV)  \\
    \hline
    13.38 & 4.24 & 5.69 & 3.41 & 0.06 & 2.00&  -2.16 & 6.06 \\
    \hline \hline
    \end{tabular}
\end{table}

\section{Appendix: Static \textit{g}-tensors under Confinement Anisotropy}
\label{app:static_g_tensors}

In the Result, we demonstrated that the driving efficiency can be affected by tuning the lateral confinement of the quantum dot. Here we explicitly list the calculated static $g$-tensors for the symmetric case and the two anisotropic geometries considered.

For the symmetric square confinement $(L_x, L_y) = (50, 50)$~nm, the numerical extraction yields a highly anisotropic but diagonally symmetric tensor:
\begin{equation}
    \mathbf{g}^{(50,50)} = \begin{pmatrix}
    0.18 & 0 & 0 \\
    0 & -0.18 & 0 \\
    0 & 0 & 21.27
    \end{pmatrix}.
\end{equation}
The dominant $z$-component ($g_{zz} \approx 21.27$) confirms the heavy-hole character.

Breaking the lateral symmetry introduces a substantial splitting between the in-plane principal values. For the geometry elongated along the $x$-direction [$(L_x, L_y) = (50, 45)$~nm], the tensor becomes:
\begin{equation}
    \mathbf{g}^{(50,45)} = \begin{pmatrix}
        0.09 & 0 & 0 \\
        0 & -0.27 & 0 \\
        0.00 & 0 & 21.27
    \end{pmatrix}.
    \label{eq:g_tensor_50_45}
\end{equation}
Conversely, rotating the confinement potential by $90^\circ$ to $(L_x, L_y) = (45, 50)$~nm inverts the anisotropy of the in-plane components:
\begin{equation}
    \mathbf{g}^{(45,50)} = \begin{pmatrix}
        0.27 & 0 & 0 \\
        0 & -0.09 & 0 \\
        0.00 & 0 & 21.27
    \end{pmatrix}.
    \label{eq:g_tensor_45_50}
\end{equation}
These results confirm the theoretical prediction that confinement anisotropy shifts the g tensor ($g_{ii} \propto \pm 3q - \alpha$).

Finally, we comment on the validity of the effective Hamiltonian used in these calculations. Our model restricts the hole dynamics to the ground-state, neglecting the coupling to higher vertical subbands via the $S$ term in the Hamiltonian. While this vertical coupling is known to introduce corrections to the static out-of-plane component $g_{zz}$~\cite{Ares2013, Watzinger2016a, Wang2024a}, these corrections have a negligible effect on the Rabi frequency in our setup. Since we operate with an in-plane magnetic field, the driving efficiency depends only on the in-plane g-factors, which are dominated by HH-LH mixing within the ground state rather than vertical subband effects.

\end{document}


\onecolumngrid
\clearpage
\begin{center}
\textbf{\large Supplemental Material for "Ge hole spin control using acoustic waves"}
\end{center}

\setcounter{equation}{0}
\setcounter{figure}{0}
\setcounter{table}{0}
\setcounter{section}{0}
\setcounter{page}{1}
\renewcommand{\theequation}{S\arabic{equation}}
\renewcommand{\thefigure}{S\arabic{figure}}
\renewcommand{\thetable}{S\arabic{table}}
\renewcommand{\thesection}{S\Roman{section}}

\section{Analytical Derivation of the Effective {$g$}-tensor}

To interpret the qubit response mechanisms, we derive an effective $2 \times 2$ Hamiltonian acting on the qubit subspace using a Schrieffer-Wolff (SW) transformation. This perturbative approach allows us to explicitly relate the macroscopic $g$-tensor components to the microscopic heavy-hole (HH) and light-hole (LH) mixing terms~\cite{Winkler2003}.

\paragraph{Schrieffer-Wolff Transformation.}
The system is described by the $4 \times 4$ Hamiltonian $H = H_{LK} + H_{BP} + H_Z$ in the $J=3/2$ basis. Due to strong quantum confinement, the heavy-hole (HH, $m_J=\pm 3/2$) and light-hole (LH, $m_J=\pm 1/2$) bands are separated by a splitting $\Delta_{LH} \equiv E_{\text{LH}} - E_{\text{HH}}$. We define the qubit subspace spanned by the HH states $\{|h\rangle, |h'\rangle\}$ and the excited subspace spanned by the LH states $\{|l\rangle\, |l'\rangle\}$.

We treat the off-diagonal coupling between these subspaces as a perturbation. To second order in perturbation theory, the effective Hamiltonian element connecting two qubit states $|h\rangle$ and $|h'\rangle$ is given by:
\begin{equation}
    \delta H_{hh'}^{(2)} \approx \sum_{l} \frac{1}{E_{h} - E_{l}} \langle h | H_{\text{mix}} | l \rangle \langle l | H_{Z} | h' \rangle + \text{h.c.},
    \label{eq:SW_component_form}
\end{equation}
where the summation runs over the LH states $\{|l\rangle\, |l'\rangle\}$. The perturbation is decomposed into the field-independent mixing term $H_{\text{mix}}$ (arising from $H_{LK}$ and strain) and the Zeeman term $H_Z$. This formula captures the dominant cross-terms where the magnetic field couples a HH state to an LH state, which is then coupled to the HH subspace via structural or strain-induced mixing.

\paragraph{Evaluation of Matrix Elements.}
In the basis ordered as $\{+\frac{3}{2}, -\frac{3}{2}, +\frac{1}{2}, -\frac{1}{2}\}$, the relevant block structure of the Luttinger-Kohn and strain Hamiltonian couples HH and LH states via the $R$ and $S$ terms:
\begin{equation}
H_{LK}+H_{BP} = 
\begin{pmatrix}
H_{\text{HH}} & H_{\text{mix}}^\dagger \\
H_{\text{mix}} & H_{\text{LH}}
\end{pmatrix}, 
\quad \text{where} \quad
H_{\text{mix}} = \begin{pmatrix} -S^\dagger & R^\dagger \\ R^\dagger & S \end{pmatrix}.
\end{equation}
Here, $R$ and $S$ represent the standard $k \cdot p$ and strain matrix elements connecting HH and LH bands.

The Zeeman term $H_Z$ allows transitions with $\Delta m_J = \pm 1$. The relevant matrix elements coupling the subspaces for an in-plane magnetic field are:
\begin{equation}
    \langle \pm\tfrac{1}{2}| H_Z |\pm\tfrac{3}{2}\rangle = \sqrt{3}\mu_B \left(\kappa + \tfrac{7}{4}q\right)(B_x \mp i B_y).
    \label{eq:HZ_HH_LH_inplane}
\end{equation}
By substituting Eq.~(\ref{eq:HZ_HH_LH_inplane}) and the explicit forms of $R$ and $S$ into the master equation [Eq.~(\ref{eq:SW_component_form})], we calculate the effective Zeeman terms.

Matching the resulting matrix elements to the definition $H_{\text{eff}} = \frac{1}{2}\mu_B \boldsymbol{\sigma} \cdot \mathbf{g} \cdot \mathbf{B}$, we obtain the analytical expressions for the in-plane $g$-factors:
\begin{subequations}
\begin{align}
    g_{xx} &= 3q - \frac{4\sqrt{3}}{\Delta_{LH}} \mathrm{Re}(R) \left(\kappa + \tfrac{7}{4}q\right), \label{eq:gxx_R_result} \\
    g_{yy} &= 3q + \frac{4\sqrt{3}}{\Delta_{LH}} \mathrm{Re}(R) \left(\kappa + \tfrac{7}{4}q\right). \label{eq:gyy_R_result}
\end{align}
\end{subequations}
Similarly, the shear-strain coupling (contained in $S$) generates the off-diagonal component $g_{zx}$:
\begin{equation}
    g_{zx} = \frac{4\sqrt{3}}{\Delta_{LH}} \mathrm{Re}(S) \left(\kappa + \tfrac{7}{4}q\right).
    \label{eq:gzx_S_result}
\end{equation}
Crucially, the SAW strain field $\boldsymbol{\epsilon}(t)$ modulates both $R$ (via $\epsilon_{xx}-\epsilon_{yy}$) and $S$ (via $\epsilon_{xz}$). Consequently, the qubit driving arises from the time-dependent modulation of the entire $g$-tensor. While $g_{zx}$ enables driving via an out-of-plane field component, the modulation of diagonal components $g_{xx}$ and $g_{yy}$ provides an additional driving mechanism when the magnetic field lies in the plane, as observed in our simulation.

Finally, we address the modulation of $g_{zz}$ and $g_{xz}$, which are not captured by this mechanism. The out-of-plane Zeeman interaction ($\propto J_z, J_z^3$) is strictly diagonal in the angular momentum basis and thus cannot directly couple the HH ($m_J = \pm 3/2$) and LH ($m_J = \pm 1/2$) subspaces. Consequently, the cross-term $\langle h | H_{\text{mix}} | l \rangle \langle l | H_{Z}(B_z) | h' \rangle$ vanishes. This implies that $B_z$ contributes to effective $g$-tensor modulation only via higher-order perturbations, which are negligible in the thin-dot limit ($\Delta_{LH} \gg H_{\text{mix}}, H_Z$).

\section{Differential Extraction Protocol}
The effective Zeeman Hamiltonian for the qubit is generally written as:
\begin{equation}
    H_{Z}^{\text{eff}} = \frac{1}{2} \mu_B \sum_{i,j} \sigma_i g_{ij} B_j,
\end{equation}
where $i, j \in \{x, y, z\}$, $\sigma_i$ are the Pauli matrices defined in the qubit basis, and $g_{ij}$ are the elements of the generalized $g$-tensor.

To extract the components $g_{ij}$ numerically, we compute the effective Hamiltonian $H_{\text{eff}}(\mathbf{B})$ at a reference field $\mathbf{B}$ and at a perturbed field $\mathbf{B} + \delta B \hat{e}_j$ (with $\delta B = 1~\mu\text{T}$). The partial derivative of the Hamiltonian with respect to the magnetic field component $B_j$ is obtained via finite differences:
\begin{equation}
    \frac{\partial H_{\text{eff}}}{\partial B_j} \approx \frac{H_{\text{eff}}(\mathbf{B} + \delta B \hat{e}_j) - H_{\text{eff}}(\mathbf{B})}{\delta B}.
\end{equation}
Using the trace orthogonality property of Pauli matrices, $\text{Tr}(\sigma_i \sigma_k) = 2\delta_{ik}$, we extract individual tensor components via projection:
\begin{equation}
    g_{ij} = \frac{1}{\mu_B} \text{Tr}\left( \sigma_i \frac{\partial H_{\text{eff}}}{\partial B_j} \right).
\end{equation}
This method allows us to capture the full, generally asymmetric, interaction matrix that describes how the magnetic field couples to the qubit spin components.

Using the material parameters for Germanium and the square comfinement defined in the main text, we first evaluate the static $g$-tensor in the absence of SAW strain ($U=0$). The numerical extraction yields a highly anisotropic, diagonal tensor:
\begin{equation}
    \mathbf{g}^{(0)} \approx \begin{pmatrix}
    0.18 & 0 & 0 \\
    0 & -0.18 & 0 \\
    0 & 0 & 21.27
    \end{pmatrix}.
\end{equation}
The dominant $z$-component ($g_{zz} \approx 21.27$) confirms the heavy-hole character of the ground states, consistent with the spin physics expected in compressively strained Ge quantum wells. The small but finite in-plane components ($|g_{xx}| \approx |g_{yy}| \approx 0.18$) arise from the finite light-hole mixing induced by the quantum confinement and the cubic symmetry terms in the Luttinger Hamiltonian. Note that coupling to higher light-hole subbands introduces corrections to the static out-of-plane component $g_{zz}$ as discussed in the Appendix~\cite{Ares2013, Watzinger2016a}.

\section{Analytical Derivation of the SAW-Driven Rabi Frequency}

In this section, we provide a rigorous derivation of the Rabi frequency under the influence of a surface acoustic wave (SAW), accounting for the elliptical polarization of the effective driving field. We demonstrate that the driving efficiency is determined by the interplay between the strain-induced effective driving field components and confirm the linear dependence on the SAW amplitude $U$.

The time-dependent interaction Hamiltonian is given by:
\begin{equation}
    H_{\text{drive}}(t) = \frac{1}{2} \mu_B \boldsymbol{\sigma} \cdot \delta\mathbf{g}(t) \cdot \mathbf{B}.
    \label{eq:H_drive_supp}
\end{equation}
Unlike standard electric dipole spin resonance (EDSR) where the driving typically oscillates linearly, the SAW strain field in a piezoelectric heterostructure involves coupled longitudinal and transverse modes with a material-dependent phase difference $\delta$. Consequently, the time-dependent $g$-tensor modulation can be expressed as a sum of in-phase and quadrature components:
\begin{equation}
    \delta\mathbf{g}(t) = U \left[ \mathcal{G}_1 \cos(\omega t) + \mathcal{G}_2 \sin(\omega t) \right],
\end{equation}
where $U$ is the acoustic displacement amplitude, and $\mathcal{G}_1, \mathcal{G}_2$ are time-independent tensors derived from the strain profile and the spin-strain coupling constants. The term $\mathcal{G}_2$ arises explicitly due to the phase difference $\delta \approx 158^\circ$ between the strain components $\epsilon_{xx}$ and $\epsilon_{xz}$.

This modulation results in an effective driving vector $\mathbf{B_{eff}} \equiv \delta\mathbf{g}(t) \cdot \mathbf{B}$ that traces an elliptical trajectory in real space.

\paragraph{Chiral Decomposition and Geometric Interpretation}

To extract the Rabi frequency, we analyze the projection of this driving vector onto the plane transverse to the static quantization axis $\mathbf{n} \parallel \mathbf{B}$. As visualized in Fig.~4 of the main text, the transverse vector $\mathbf{B_{eff}}_{\perp}(t)$ traces an ellipse defined by a semi-major axis $R_{\text{major}}$ and a semi-minor axis $R_{\text{minor}}$.

Any elliptical motion can be mathematically decomposed into two circular components rotating in opposite directions: a co-rotating component (radius $R_{\text{major}}+R_{\text{minor}}$) and a counter-rotating component (radius $R_{\text{major}}-R_{\text{minor}}$). Within the Rotating Wave Approximation (RWA), only the co-rotating component which matches the helicity of the hole spin Larmor precession contributes to the spin transition.

The magnitude of this effective driving vector is given by the superposition of the major and minor axes, governed by the chirality of the ellipse relative to the spin precession:
\begin{equation}
    |\mathbf{B}_{\text{drive}}| = \frac{1}{2} (R_{\text{major}} \pm R_{\text{minor}}).
    \label{beff}
\end{equation}

The sign choice in Eq.~(\ref{beff}) is determined by the relative chirality between the driving vector and the spin Larmor precession. When the driving vector co-rotates with the Larmor precession, the dominant elliptical component $R_{\text{major}} + R_{\text{minor}}$ term results in efficient resonant coupling, which corresponds to the maximal Rabi frequency observed at $\theta \approx 330^\circ$. Conversely, when the vector rotates in the opposite direction (counter-rotating case), the dominant elliptical component is off-resonant, and the spin is driven solely by the weak residual difference between the axes ($R_{\text{major}} - R_{\text{minor}}$), explaining the strong suppression observed at $\theta \approx 30^\circ$.

Since both $R_{\text{major}}$ and $R_{\text{minor}}$ scale linearly with the displacement amplitude $U$, the Rabi frequency can be expressed in a compact geometric form:
\begin{equation}
    \Omega_R = \frac{\mu_B}{2\hbar} |\mathbf{B}_{\text{drive}}| = \frac{\mu_B}{4\hbar} (R_{\text{major}} \pm R_{\text{minor}}).
    \label{eq:Rabi_geometric}
\end{equation}
This formula provides a direct link between the orbital geometry of the effective vector and the qubit control speed.

\paragraph{Validation against Numerical Simulation}

To validate this geometric analytical model, we compare the calculated $\Omega_R$ from Eq.~(\ref{eq:Rabi_geometric}) with the results obtained from full numerical integration of the time-dependent Schrödinger equation (using QuTiP).

\begin{figure}[h!]
    \centering
    \includegraphics[width=0.7\columnwidth]{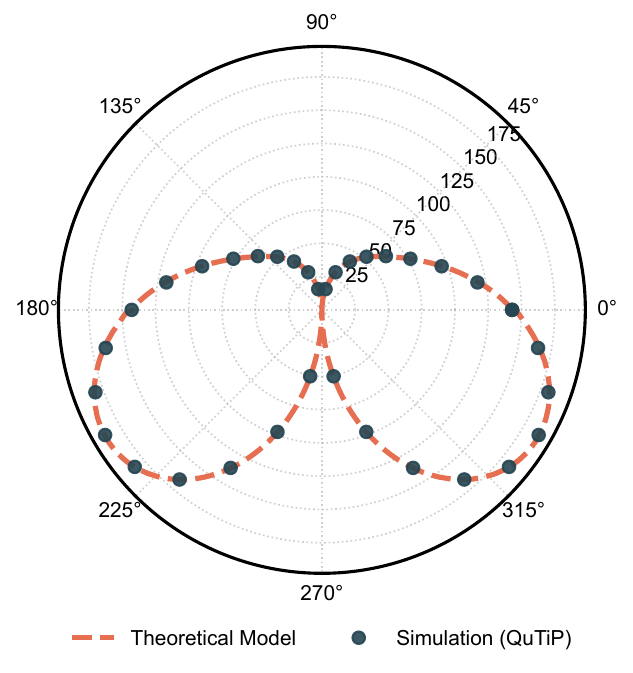}
    \caption{\textbf{Verification of Geometric Anisotropy.} Comparison of the Rabi frequency anisotropy calculated from the full numerical simulation (blue dots) and the analytical elliptical geometry model (dashed orange line). The perfect agreement confirms that the chiral selection rule—based on the sum or difference of the ellipse axes—accurately describes the driving mechanism.}
    \label{fig:anisotropy}
\end{figure}

As shown in Fig.~\ref{fig:anisotropy}, the analytical prediction (dashed orange line) shows excellent agreement with the numerical simulation (blue dots) across all magnetic field angles.

\begin{figure}[h!]
    \centering
    \includegraphics[width=0.7\columnwidth]{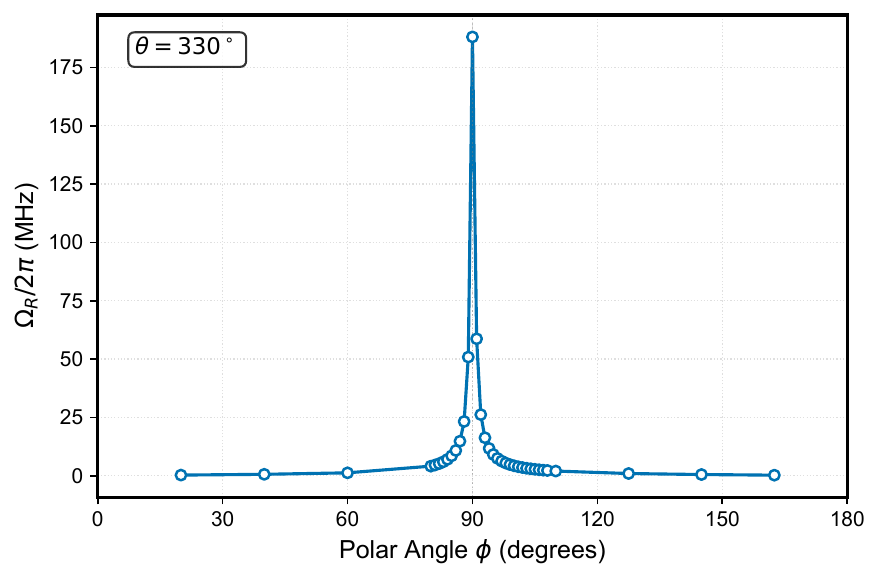}
   \caption{Simulated SAW-driven Rabi frequency \(\Omega_R/2\pi\) versus polar angle \(\theta\) for fixed azimuth \(\phi=330^\circ\). A single peak appears near \(\theta\approx 90^\circ\), while \(\Omega_R\) is strongly suppressed away from the \(xy\) plane due to the large \(g_{zz}\).}
    \label{fig:anisotropy_3d}
\end{figure}

\paragraph{Rabi frequency vs.\ polar angle at fixed \(\phi\).}
We scan the polar angle \(\theta\) at fixed azimuth \(\phi=330^\circ\). As shown in Fig.~\ref{fig:anisotropy_3d}, owing to the strong \(g\)-tensor anisotropy (large \(g_{zz}\)), the resonant field magnitude \(|\mathbf{B}_{\rm res}|\) decreases rapidly once \(\mathbf{B}\) tilts away from the \(xy\) plane. Since the SAW-induced drive \(\mathbf{B_{eff}}=\delta\mathbf{g}(t)\!\cdot\!\mathbf{B}\) scales with \(|\mathbf{B}_{\rm res}|\), the Rabi frequency is strongly suppressed for out-of-plane tilts, leaving an appreciable response only near \(\theta\simeq 90^\circ\) and producing the single peak in the \(\theta\) sweep.
\section{Rabi freq under confinement anisotropy}
\label{sec:opt_geometry}
\paragraph{Optimization of driving efficiency via confinement anisotropy}
The power dependence (slope) of the Rabi frequency reported in the main text can be related to the static magnetic field required to fulfill the resonance condition. For an in-plane magnetic field and neglecting small in-plane off-diagonal tensor elements, the effective Zeeman splitting is
\begin{equation}
E_Z(\phi)=\mu_B B\, g_{\rm eff}(\phi),
\qquad
g_{\rm eff}(\phi)=\sqrt{g_{xx}^2\cos^2\phi+g_{yy}^2\sin^2\phi}.
\end{equation}
At a fixed driving frequency $f$, the resonance field is then
\begin{equation}
B_{\rm res}(\phi)=\frac{h f}{\mu_B\, g_{\rm eff}(\phi)}.
\label{eq:Bres_geff}
\end{equation}
In the SAW-driven $g$-tensor modulation scheme considered here, for a fixed SAW amplitude and fixed chirality (i.e., fixed co-rotating component selected under the RWA), the Rabi frequency scales linearly with the resonant static field magnitude, so maximizing the driving efficiency is equivalent to minimizing $g_{\rm eff}(\phi)$.

Breaking lateral symmetry ($L_x\neq L_y$) yields $\langle k_x^2-k_y^2\rangle\neq 0$, activating a finite real part of the kinetic HH--LH mixing term $R$. Using Eqs.~(\ref{eq:gxx_R_result})--(\ref{eq:gyy_R_result}), we write
\begin{equation}
g_{xx}=3q-\alpha,\qquad g_{yy}=-3q-\alpha,
\label{eq:gxxgyy_alpha}
\end{equation}
with
\begin{equation}
\alpha\equiv
\frac{4\sqrt{3}}{\Delta_{LH}}\,
\mathrm{Re}(R)\left(\kappa+\frac{7}{4}q\right).
\label{eq:alpha_def}
\end{equation}

Within the hard-wall ground-state approximation used throughout this work,
$\langle k_i^2\rangle=(\pi/a_i)^2$, hence
\begin{equation}
\mathrm{Re}(R)= -\frac{\hbar^2\sqrt{3}}{2m_0}\gamma_2\,\pi^2
\left(\frac{1}{L_x^2}-\frac{1}{L_y^2}\right).
\label{eq:ReR_hardwala_exact}
\end{equation}
Combining Eqs.~(\ref{eq:alpha_def}) and (\ref{eq:ReR_hardwala_exact}) gives the desired closed-form relation between the in-plane $g$-factor shift and the lateral confinement dimensions:
\begin{equation}
\alpha=
-\frac{6\pi^2\hbar^2}{m_0}\;
\frac{\gamma_2}{\Delta_{LH}}\;
\left(\kappa+\frac{7}{4}q\right)
\left(\frac{1}{L_x^2}-\frac{1}{L_y^2}\right).
\label{eq:alpha_LxLy_exact}
\end{equation}

Using Eq.~(\ref{eq:Bres_geff}) and the parametrization in Eq.~(\ref{eq:gxxgyy_alpha}),
the geometry that maximizes the driving efficiency at a given in-plane angle $\phi$ is obtained by minimizing
\begin{equation}
g_{\rm eff}^2(\phi)
=(3q-\alpha)^2\cos^2\phi+(3q+\alpha)^2\sin^2\phi
\end{equation}
with respect to $\alpha$. This yields the optimal condition
\begin{equation}
\alpha_\star(\phi)=3q\cos(2\phi).
\label{eq:alpha_opt}
\end{equation}
Equivalently, the optimal in-plane principal values become
\begin{equation}
g_{xx}^\star(\phi)=3q-\alpha_\star(\phi)=6q\sin^2\phi,
\qquad
g_{yy}^\star(\phi)=-3q-\alpha_\star(\phi)=-6q\cos^2\phi,
\label{eq:gstar}
\end{equation}
which minimizes $g_{\rm eff}(\phi)$ and thus maximizes $B_{\rm res}(\phi)$ and $\Omega_R(\phi)$ for that specific field angle.

Finally, inserting Eq.~(\ref{eq:alpha_opt}) into the closed-form expression
(\ref{eq:alpha_LxLy_exact}) gives an explicit design equation for the confinement anisotropy:
\begin{equation}
\left(\frac{1}{L_x^2}-\frac{1}{L_y^2}\right)
=
-\frac{m_0\,\Delta_{LH}}{2\pi^2\hbar^2}\;
\frac{q\cos(2\phi)}{\gamma_2\left(\kappa+\frac{7}{4}q\right)}.
\label{eq:design_rule_LxLy}
\end{equation}

Equation~(\ref{eq:design_rule_LxLy}) provides a direct geometric guideline: for any chosen in-plane operation angle $\phi$, one can tune the lateral aspect ratio $(L_x,L_y)$ such that the $R$-induced shift of $(g_{xx},g_{yy})$ minimizes $g_{\rm eff}(\phi)$, thereby maximizing the resonant field and the resulting SAW-driven Rabi frequency.
Equation~(\ref{eq:design_rule_LxLy}) provides a direct geometric guideline: for any chosen in-plane operation angle $\phi$, one can tune the lateral aspect ratio $(L_x,L_y)$ such that the $R$-induced shift of $(g_{xx},g_{yy})$ minimizes $g_{\rm eff}(\phi)$, thereby maximizing the resonant field and the resulting SAW-driven Rabi frequency.

It is worth noting, however, that while the theoretical optimum suggests minimizing $g_{\rm eff}$ to zero, practical experimental constraints impose a lower bound on the attainable $g$-factor. As $g_{\rm eff} \to 0$, the required resonant field $B_{\rm res}$ diverges, quickly exceeding the range of standard vector magnets. Furthermore, operating at extremely high magnetic fields may introduce unwanted coupling to excited orbital states or compromise qubit coherence due to enhanced sensitivity to parameter fluctuations. Therefore, the optimal geometric design represents a trade-off: the confinement anisotropy should be tuned to reduce $g_{\rm eff}$ sufficiently to enhance the driving speed, while ensuring that the required $B_{\rm res}$ remains within a technically feasible and controllable regime.
\paragraph{Interplay between elliptical driving and confinement-induced $g$-tensor anisotropy }
The optimization analysis above was carried out at a fixed in-plane operation angle $\phi$, such that the driving efficiency could be discussed solely in terms of the resonance field $B_{\rm res}(\phi)$ through $g_{\rm eff}(\phi)$. In the general case of an in-plane angular sweep, however, the observed Rabi-frequency anisotropy results from two concurrent mechanisms: (i) the chiral, elliptically polarized SAW driving which selects either the co-rotating or counter-rotating component depending on the relative chirality, and (ii) the confinement-induced shift of the in-plane $g$-tensor principal values through the kinetic mixing term $R$, which reshapes $g_{\rm eff}(\phi)$ and therefore the resonant field. Because these two effects have distinct angular dependences, their superposition can yield nontrivial line shapes that cannot be captured by a single-parameter optimization at fixed $\phi$.

\begin{figure}[h!]
    \centering
    \includegraphics[width=0.75\columnwidth]{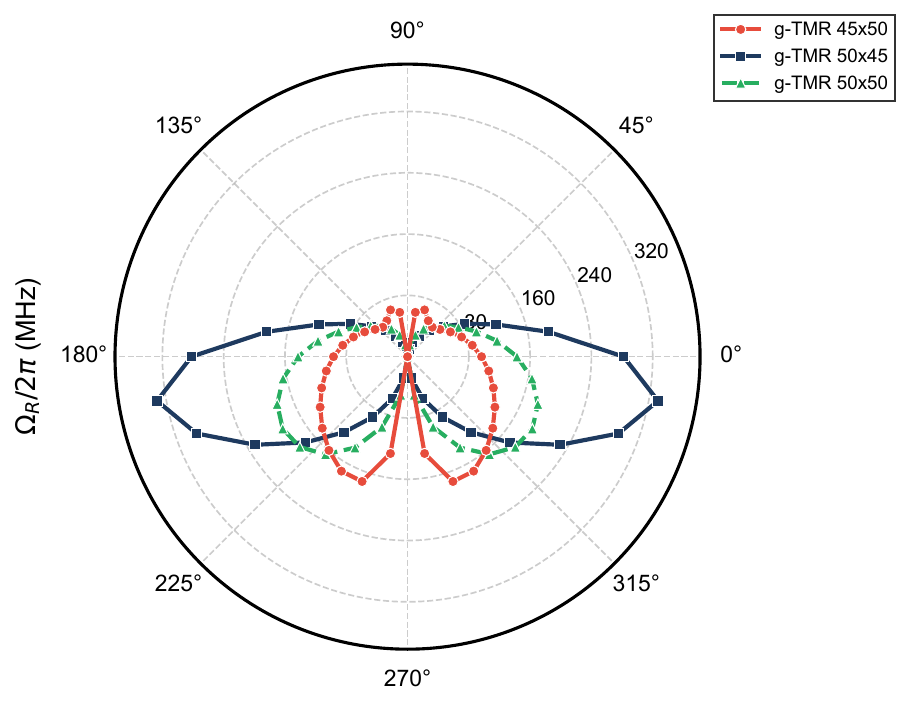}
    \caption{\textbf{Angular dependence of SAW-driven Rabi frequency under confinement anisotropy.}
    Polar plot of $\Omega_R/2\pi$ for an in-plane sweep of the static magnetic-field direction, comparing three lateral confinement geometries. The symmetric case  $(L_x,L_y)=(50~\mathrm{nm},50~\mathrm{nm})$ (green triangles) serves as a reference. Breaking the lateral symmetry by swapping the dimensions to $(45~\mathrm{nm},50~\mathrm{nm})$ (red circles) and $(50~\mathrm{nm},45~\mathrm{nm})$ (blue squares) results in distinct angular line shapes, demonstrating how geometric anisotropy can be used to engineer the direction of maximum driving efficiency.}
    \label{fig:anisotropy_confine}
\end{figure}

In particular, Fig.~\ref{fig:anisotropy_confine} compares the angular response of a symmetric dot $(L_x,L_y)=(50~\mathrm{nm},50~\mathrm{nm})$ against two anisotropic configurations related by swapping the lateral dimensions $(50~\mathrm{nm},45~\mathrm{nm})$ and $(45~\mathrm{nm},50~\mathrm{nm})$. The symmetric case (green trace) serves as a baseline, exhibiting a pattern influenced solely by the elliptically polarized SAW driving. Introducing confinement anisotropy adds a second effect: for $(45,50)$, the Rabi frequency is suppressed along the x-axis, whereas for $(50,45)$, it is enhanced in that direction. These trends highlight that, when sweeping the field direction, the optimal angle is not universal but is set by the interplay between the elliptically polarized driving field and the confinement-modified $g$-tensor anisotropy.

\bibliography{SAW_paper_ref}